\documentclass[aps,prb,twocolumn,floats]{revtex4-1}
\usepackage{graphicx}
\usepackage{amsmath}
\usepackage{amssymb}
\usepackage{bm}
\usepackage{mathtools}
\usepackage[normalem]{ulem}
\usepackage{color}
\usepackage{tikz}
\usepackage{textcomp}
\usepackage[colorlinks,linkcolor=red,citecolor=blue,filecolor=magenta,
urlcolor=cyan]{hyperref}

\begin{document}
\title{Theory of Optimal transport and the structure of Many-Body states}
\author{S. R. Hassan}
\email{shassan@imsc.res.in}
\affiliation{The Institute of Mathematical Sciences, C.I.T. Campus, Chennai 600 
113, India}
\affiliation{Homi Bhabha National Institute, Training School Complex, Anushakti Nagar, Mumbai 400094, India}
\author{Ankita Chakrabarti}
\email{ankitac@imsc.res.in}
\affiliation{The Institute of Mathematical Sciences, C.I.T. Campus, Chennai 600 
113, India}
\affiliation{Homi Bhabha National Institute, Training School Complex, Anushakti Nagar, Mumbai 400094, India}
\author{R.Shankar}
\email{shankar@imsc.res.in}
\affiliation{The Institute of Mathematical Sciences, C.I.T. Campus, Chennai 600 
113, India}
\affiliation{Homi Bhabha National Institute, Training School Complex, Anushakti Nagar, Mumbai 400094, India}

\date{\today}
\begin{abstract}
There has been much work in the recent past in developing the idea of quantum
geometry to characterize and understand the structure of many-particle states.
For mean-field states, the quantum geometry has been defined and analysed in
terms of the quantum distances between two points in the space of single
particle spectral parameters (the Brillioun zone for periodic systems) and the
geometric phase associated with any loop in this space. These definitions are
in terms of single-particle wavefunctions. In recent work, we had proposed a
formalism to define quantum distances between two points in the spectral
parameter space for any correlated many-body state. In this paper we argue
that, for correlated states, the application of the theory of optimal transport to
analyse the geometry is a powerful approach. This technique enables us to
define  geometric quantities which are averaged over the entire spectral
parameter space.  We present explicit results for a well studied model, the one
dimensional $t-V$ model, which exhibits a metal-insulator transition, as
evidence for our hypothesis.

\end{abstract}

\maketitle
\section{Introduction}
\label{intro}
Many-particle wavefunctions are complex functions of a large number of
variables. Developing good techniques to visualize them and characterize their
structure can contribute to the understanding of a large number of physical
systems. In recent work \cite{Paper1,Paper2}, we have proposed a
mathematically consistent definition of distances between two points in the
spectral parameter space for a general many-fermion state. Our definition
generalises the previous definition which was in terms of the single-particle
wavefunctions and hence was only valid for mean-field states. 

We had implemented our definition for the well studied 1 dimensional $t-V$
model, which exhibits a metal-insulator transition. Our work was motivated by
the seminal paper of Walter Kohn \cite{Walterkohn} and work that followed it
\cite{RestaSorella,SCPMR,SouzaMartin,MarrazzoResta,resta2011}. 
Kohn \cite{Walterkohn} proposed that the structure of the ground state
alone could distinguish between a metallic and an insulating system. He argued
that the key feature of insulating states which characterises it, is the fact that they are
insensitive to changes in the boundary conditions. He proposed a general form
of such wavefunctions and hypothesised that the ground state wave function of
all insulating states were of that form. These wave functions are sharply
localized about a set of regions in the configuration space.

While this hypothesis is conceptually appealing, it is diffcult to
implement in practice. For example, even if we were given the ground state wave
function (say by numerical diagonalization), checking if it is of the form
hypothesized by Kohn is a very difficult problem. Hence, the work following up 
on Kohn's idea \cite{RestaSorella,SCPMR,SouzaMartin,MarrazzoResta,resta2011}, 
concentrated on finding simpler ways to characterise the localization of the 
ground state wavefunction in the configuration space.

Any many-body state is characterised by its static correlation functions.
Thus, the Kohn's proposal can be rephrased to say that metals and insulators
can be distinguished by certain static ground state correlation functions.
Resta and Sorella identified such correlations. They proposed
\cite{RestaSorella} that the localization tensor, which is the second moment
of the pair correlation function is such a quantity.  They showed that it is
finite in the insulating state and diverges in the metallic state.

The interesting aspect of this approach is that it can be related to concepts
of quantum geometry. For mean field states describing band insulators, the
localisation tensor can be shown to be the integral of the quantum metric over
the Brillioun zone \cite{SCPMR}. The quantum metric is defined in terms of the
single-particle Bloch wavefunctions in the standard way \cite{Bures,resta2011}.
For correlated states, Souza et. al. showed \cite{SouzaMartin} that it
can be written as average over the space of twisted boundary conditions of a
metric defined on the manifold of ground states of the system with twisted
boundary conditions. For mean field states, this expression reduces to the
standard one described above. The body of work discussed above motivates the
rephrasing of Kohn's words ``organisation of the electrons in the ground
state'' as ``quantum geometric structure of the ground state ''.

The localization tensor is a coarse grained description of the quantum geometric
structure of the ground state. It only describes averages, spatial averages of
the pair correlation function, or equivantly, Brillioun zone averages of the
quantum metric.  Recent work \cite{MarrazzoResta} attempts to generalize the
concept locally in space, with potential application to inhomogeneous systems.
However, even for translationally invariant systems, there is more detailed
physical information in the quantum geometry than averaged quantities.

Motivated by the above discussion and our recent work \cite{Paper1,Paper2}, in
this paper we attempt to develop a formalism to bring out the detailed geometry
of translationally invariant, correlated states. Our previous numerical results
\cite{Paper1,Paper2} indicate that while it is possible to give a
mathematically consistent definition of quantum distances between two points in
the Brillioun zone (BZ), the differential metric may not exist for correlated
states. Thus, we develop the formalism in terms of quantum distances rather than
the differential quantum metric.  

For a lattice translation invariant single band model (like the $t-V$ model),
our definition of the distance between two points in the BZ can be
qualitatively thought of as a measure of the difference in the occupancies of
these two points. The concept of a distance distribution defined at every
point in the BZ, is then useful to characterize the metallicity of the state.
Intiutively, in the metallic state, we expect the occupancy of the points in
the Fermi sea and those outside it to be very different. On the other hand,
deep in the insulating regime, since we expect the kinetic energy to be
quenched, there should not be much difference between the occupancy of the
various points in the BZ.

To characterize the above behaviour using a concrete quantitative approach, we
apply the theory of optimal transport \cite{Villani,villani2003,OT_2,WBC1,Cuturi,OT_ML,OT_data}, which can
define distances between the distance distributions at any two points in the BZ
in terms of the so called Wasserstein distances. Optimal transport theory has been
first applied in condensed-matter physics in the context of the density functional theory \cite{OT_DFT,OT_DFT_2}.
The Wasserstein distance between any two distributions is the weighted average of all 
the distances over the BZ, where the weights are specified by an optimal joint probability
distribution whose marginals are given by the above distribution functions.
Based on our numerical and analytical results on the one dimensional $t-V$ model, we
conjecture that it is a useful quantity to characterize the metallic and
insulating states.  Further, using this formalism, we identify a single
distribution function on the BZ, the Wasserstein barycenter, which can sense the
metal-insulator transition. The average Wasserstein distance between the
barycenter and all the distance distributions, is identified as a single
parameter which may provide a clear distinction between the metallic and
insulating phases in the thermodynamic limit.

Our previous work \cite{Paper1,Paper2} and this one constitutes our
attempt to implement our definition of quantum distances for correlated states
in the physical context of the work of Kohn and others  
\cite{RestaSorella,SCPMR,SouzaMartin,MarrazzoResta,resta2011} towards a
quantum geometric characterization of the insulating state. All our results in
these papers are obtained by exact diagonalization of the one dimensional $t-V$
model  \cite{Yang-XXZ,Baxter,tV-bosonisation}. For our initial investigations we 
chose to concentrate on this model for the following reasons, 
(a) it is a well studied model that exhibits a metal-insulator transition, 
(b) it is a one band model. The definition of quantum
distances in terms of single particle wavefunctions yields no non-trivial
results for one band models. So it is an ideal model to study the effects of
correlations on the quantum geometry. Based on the insights obtained from this
study, we hope to report results in the future on multi-band models and in
different physical contexts.

The rest of this paper is organised as follows. In Section \ref{revpw}, we
briefly review the results obtained in our previous work.  Section \ref{revott}
reviews the optimal tansport theory in a general context. Section \ref{WQD}
describes how we apply the theory in the context of many-body states to define
the Wasserstein distance in terms of the quantum distances. We present analytic
results for the Wasserstien distance for the ground state of our model for the
extreme limits of the interaction strength and numerical results for
intermediate interaction strengths in this section.  Section \ref{WED}
describes how we apply the theory and define the Wasserstein distance in terms
of the Euclidean distances defined on the BZ.  This is followed by numerical
results.  Section \ref{WBC} discusses further application of optimal transport
theory to define the geometrical concept of Wasserstein barycenter and the
average Wasserstein distance between the barycenter and all the distance
distributions.  This is followed by results obtained by a combination of
numerical techniques and analytical results.  Finally, we summarize our
results and discuss the conclusions we draw from them in Section \ref{dandc}.

\section{Brief review of our previous work}
\label{revpw}
In a recent paper \cite{Paper1}, we had given a definition for the quantum
distance between two points in the spectral parameter space, for a general
correlated many-fermion state. By spectral parameters, we mean the parameters
that label the single particle spectrum of the system. We had shown that our
definition reduces to the standard one \cite{Bures,resta2011} in terms of the
single-particle wave-functions for mean field states. We had also shown that
our definition satisfies the basic mathematical requirements of a distance,
including the triangle inequalities. Our definition is detailed below.
For the sake of concreteness, let us consider a translationally invariant
tight-binding lattice model, where the single particle spectrum is labelled by
$(\alpha,\bf k)$ where $\alpha=1,\dots,N_B$ is the sub-lattice index and
$\mathbf k$ are the quasi-momenta taking values in the Brillioun zone. A
general many-particle state in this model can be written in the Fock basis as,
\begin{equation}
\label{psidef}
\vert\psi\rangle=
\sum_{\{n_{(\alpha,\mathbf k)}\}}\psi(\{n_{(\alpha,\mathbf k)}\})
\vert \{n_{(\alpha,\mathbf k)}\}\rangle
\end{equation}
where $\{n_{(\alpha,\mathbf k)}\}$ denotes the set of occupation numbers.

We define occupation number exchange operators, $E(\mathbf k_1,\mathbf k_2)$,
that interchange the occupation numbers of the modes, $n_{(\alpha,\mathbf
k_1)}\leftrightarrow n_{(\alpha,\mathbf k_2)}$ for all $\alpha=1,\dots,N_B$. The
quantum distance between $\mathbf k_1$ and $\mathbf k_2$, $d(\mathbf
k_1,\mathbf k_2)$, is defined as,
\begin{equation}
\label{dk1k2def}
d(\mathbf k_1,\mathbf k_2)\equiv
\sqrt{1-\vert\langle\psi\vert E(\mathbf k_1,\mathbf k_2)\vert\psi\rangle\vert^2}
\end{equation}

We have also shown \cite{Paper1} that the occupation number exchange
operators can be explictly written in terms of the fermion creation and
annihilation operators. In general, the problem reduces to computing static
correlation functions. Specifically, in the simplest case of one band models, 
it reduces to the computation of 4-point functions \cite{Paper1}.
\begin{equation}
\label{1bandd}
\left(d(\mathbf k_1,\mathbf k_2)\right)^2=
2\langle \left(n_{\mathbf k_1}-n_{\mathbf k_2} \right)^2 \rangle 
-(\langle \left(n_{\mathbf k_1}-n_{\mathbf k_2}\right)^2\rangle)^{2}
\end{equation}
Thus, in this case, the quantum distance between two points in the BZ can be 
interpreted as a measure of the difference in the occupancies of these 
two points. For multi-band systems, computation of the quantum distances
involve the computation of higher point static correlations.

Nevertheless, the matrix of above quantum distances \cite{Paper2}, $D$, for the 
ground state of any interacting system can be explicitly computed using any technique 
(exact or approximate) that can compute static ground state correlation functions. These
include quantum Monte-Carlo methods, exact diagonalization for finite systems,
bosonization and DMRG for 1-dimensional sytems, semiclassical methods,
perturbation theory, etc..

We had used the numerical exact diagonalization method to compute the distance
matrix for a finite system of spinless interacting fermions in 1-dimension, the
so called $t-V$ model \cite{Paper1}. Our results indicated that various
properties of the distance matrix were very different in the metallic and
insulating regimes of the system. This motivated us to search for other
geometric quantities, constructed from the distance matrix, which would
distinguish more sharply between the metallic and insulating regimes
\cite{Paper2}. Our results also indicated that, in general, the quantum
distances may not define a differential quantum metric. Namely, to define a differential
metric, in the thermodynamic limit we must have 
$d(\mathbf k,\mathbf k+\mathbf dk)= g(\mathbf k)^{ij}dk_idk_j$ 
for $\vert\mathbf dk\vert<<\vert\mathbf k\vert$. Our results
for the finite system \cite{Paper1} indicate that this may not be true for
interacting systems. Hence we used the methods of discrete geometry
\cite{Paper2} to study the system.  In particular, we found that the so called
Ollivier-Ricci curvature is a good geometric quantity to distinguish between
the two regimes \cite{Paper2}.  The Ollivier-Ricci curvature is defined in
terms of the so called Wasserstein distance, which is computed using the
theory of optimal transport. We also found the approximate Euclidean embedding
of the Wasserstein distance is able to distinguish the metallic and insulating
phase.

\subsection{The distance distribution functions}
\label{DDF}

Consistent with the interpretation of the quantum distances in one band models
discussed above, we found from our numerical results for the one dimensional $t-V$ model
\cite{Paper1,Paper2} that deep in the metallic regime ($V << 1$), the distances
classify the quasi momenta inside the Fermi sea and those outside it into two
different categories.  The distances between any two points both inside or
outside the Fermi sea are very small ($\sim 0$) and those between two points,
where one lies inside the Fermi sea and the other outside it, are very large
($1$).  The points inside the Fermi sea we label as $k_{in}$ and points outside
the Fermi sea we label as $k_{out}$. On the other hand, deep in the insulating
regime ($V >> 1$), the distances are rather homogenous and does not 
distinguish very much between quasi-momenta in the Fermi sea and those 
outside it.

This behaviour motivates us to analyse the distance matrix in terms of probability distributions
$\{m_i(j)\}$, defined at each point in the BZ, $i=1,\dots,L$, 
constructed from normalised distribution of distances of all the points in the BZ from the above
point $k_i$, as follows:

\begin{equation}
\label{DF}
m_i (j)\equiv \frac{ D(i,j)}{\sum_{l=1}^{L}D(i,l)}.
\end{equation}

\begin{figure}[h!]
\centering
\includegraphics[angle=0,width=0.48\textwidth]{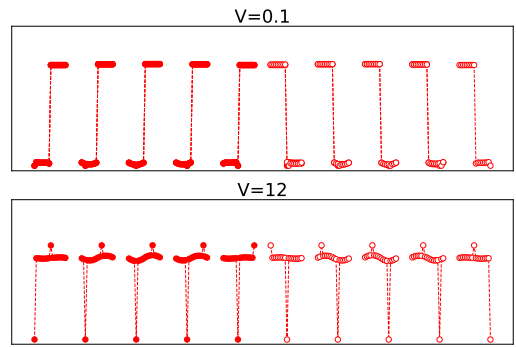}
\caption{Schematic figure depicting the distribution functions, $m_i(j)$ $(i,j=1,...,L)$, defined 
at each point in the BZ for the two regimes of interaction, for 18 sites. The first five 
distributions depicted with filled circles represent distribution functions at alternate points 
for quasi-momenta modes inside the Fermi sea, $m_{k_{in}}$. While the remaining five depicted using 
unshaded circles represent distributions for quasi momenta modes outside it, $m_{k_{out}}$. 
In deep metallic regime, at $V=0.1$, the first five distributions are completely opposite of the
next five. However, for deep insulating regime, at $V=12$, $m_i$ are almost identical
for all $k_i \in k_{in},k_{out}$, differing only at points $i,i+\frac{L}{2}$.} 
\label{PDF}
\end{figure}

Deep in the metallic regime, the above distributions $\{m_i(j)\}$ are completely opposite of each other for points 
inside Fermi sea and points outside it and exactly identical for all points inside (or outside) the Fermi sea. 
Whereas deep in insulating regime, the distribution for every point in the BZ are almost identical, 
differing only at the points $(i, i+\frac{L}{2})$ \cite{Paper1,Paper2} ( Eq.~\ref{dinfmat}). This is illustrated in the Fig. (\ref{PDF}).

In the remaining part of this paper, we elaborate on how we analyse the geometry of the above 
distance distribution functions using the technique of optimal transport.
In particular we study the construction of Wasserstein distances and investigate the 
underlying optimal transport theory in details in the context of quantum geometry of the correlated states. 
The geometrical observables thus obtained are found to 
sharply characterise the phases of the system.

\section{Theory of optimal transport and the Wasserstein distances}
\label{OT}
\subsection{Review of the theory of optimal transport}
\label{revott}
In this section, we will review the theory of optimal 
transport, focussing on the context we will be applying it to, namely
quantum geometry of the states of a 1-dimensional model of interacting
fermions.

\subsubsection{The Wasserstein distance}

The theory of optimal transport defines distances between probability distribution functions \cite{Villani,villani2003}
and thus allows us to compare a set of distance distribution functions quantitatively.

Optimal transport gives a definition of distance between two probability distribution functions (PDFs)
$m_i$ and $m_j$, $W_{p}(m_i,m_j)$, as follows:\\
\begin{equation}
W_{p}^{(p)}(m_i,m_j) \equiv \inf_{\pi} \sum_{k,l}(\tilde{d}(k,l))^{p}\pi_{ij} (k, l),
\label{Cdef}
\end{equation}
where $k,l$ in the above sum runs over the domain of the PDFS, $p \in [1, \infty)$
and $\pi_{ij}(k,l)$ are joint probability distributions whose marginals are $m_i$ and $m_j$,
\begin{equation}
\label{constraints}
\sum_l\pi_{ij}(k,l)=m_i(k),~~\sum_k\pi_{ij}(k,l)=m_j(l).
\end{equation}
The $p$th root of the above optimised function $W_{p}(m_i,m_j)$ satisfies all the axioms of a distance function only
when $\tilde{d}(k,l)$ is a valid distance function and satisfies all the properties of a metric.

Physically, $\pi_{ij}(k,l)$ is usually interpreted as different ways to 
transport material such that the distribution function $m_i$ is transformed to the 
distribution function $m_j$. However, $(\tilde{d}(k,l))^{p}$ in a more general scenario, is
called the cost function of the transport where it is not restricted to be positive powers 
of a valid distance and is interpreted as the cost paid for above transfer. 
Whereas the above optimised sum $W_{p}^{(p)}(m_i,m_j)$ is defined as the minimal cost of 
transforming $m_i$ to $m_j$. The central concept of above transport problem involves finding an optimal 
joint distribution function $\pi_{ij}^{*}(k,l)$ such that the sum defined on the RHS of Eq. \ref{Cdef} is minimum.

Choosing $p=2$ in Eq.~\ref{Cdef}, we define squared Wasserstein distances $W^{(2)}(m_i,m_j)$ between any 
two PDFs $m_i$ and $m_j$ as follows,
\begin{equation}
\label{Wdef_gen}
W^{(2)}(m_i, m_j) \equiv \inf_{\pi} \sum_{k,l} (\tilde{d}(k,l))^{2}\pi_{ij} (k, l).
\end{equation}
Here $\pi_{ij}$ satisfies the constraints given by Eq.~(\ref{constraints}).

We consider the distributions, $m_i$ and $m_j$, to be the distance distributions defined 
in Section \ref{DDF} at any two points $k_i$ and $k_j$ on the BZ.
While $\tilde{d}(k,l)$ can be any valid distance defined between the points in the BZ,
we have studied the quantum distances and the Euclidean distances on the BZ (detailed later).
The corresponding squared Wasserstein distance between the 
distance distributions, given by the above sum in Eq.~(\ref{Wdef_gen}) then gives the weighted 
average of all the squared distances between any two points in the BZ,  
with the corresponding weights given by the optimal joint 
probability distribution $\pi^{*}_{ij}$.
The Wasserstein distance can be computed numerically using the standard 
techniques of linear programming \cite{OR_LP} by minimization of the linear function of $\pi_{ij}$, 
defined in Eq.~(\ref{Wdef_gen}), subject to linear constraints specified by Eq.~(\ref{constraints}).

\subsection{The Wasserstien distance obtained from the quantum distance}
\label{WQD}
In this section we study the geometry of the distance distributions (Sec. \ref{DDF}) in terms of the Wasserstein
distance obtained by choosing the square of the quantum distances (defined in Eq.~(\ref{dk1k2def})),
as the cost function of the optimal transport problem. This is done by substituting these distances in Eq.~(\ref{Wdef_gen}). 
We first look at physical interpretation of above distances in terms of quantities in the Hilbert space, since our
definition is derived completely from the quantum distances. Finally, we close 
this section by discussing the results obtained from analytical calculations at extreme
limits of interaction and from the numerical linear programming solutions for intermediate
interaction values, using the distance matrices given by exact diagonalization.

We define squared Wasserstein distances between two distance distributions $m_i$ and $m_j$ in terms of
the matrix of quantum distances $D$ as follows,
\begin{equation}
W^{(2)}(m_i, m_j) \equiv \inf_{\pi} \sum_{k,l=1} ^{L} (D(k,l))^{2}\pi_{ij} (k, l).
\label{Wdef}
\end{equation}
Where $\pi_{ij}$ satisfies the constraints given by Eq. \ref{constraints}.
The above distance is closely related to the underlying geometry 
of the discrete metric space under investigation
and is intricately related to the intrinsic curvature \cite{OLLIVIER2009,OC2,ollivier:hal-00858008}, as discussed
in details in our previous work \cite{Paper2}.

\subsubsection{Physical interpretation}
\label{PI}
In our context, namely the quantum geometry of many-fermion states,
we have no concept of ``transport". We are studying the kinematics
of many-fermion states and hence there is no time evolution. We are only
analysing static correlation functions. 

Thus, the physical interpretation of our application of the 
theory of optimal transport is quite different from the standard one discussed 
above. It is as detailed below.

The  distance matrix defined in equation (\ref{dk1k2def}) can be written as,
\begin{eqnarray}
(D(i,j))^{2}&\equiv&1-{\rm Tr}\left(\hat\rho(i,j)\hat\rho_0\right)\\
\hat\rho_0&\equiv&\vert\psi\rangle\langle\psi\vert\\
\hat\rho(i,j)&\equiv&E(i,j)\vert\psi\rangle\langle\psi\vert E(i,j)
\end{eqnarray}
Thus, the distance matrix is defined in the subspace spanned by the states
$E(i,j)\vert\psi\rangle\equiv\vert i,j\rangle$. We will call this subspace 
the quantum distance Hilbert space, $QDH$.

The minimisation of the sum defined on the RHS of \\
Eq. \ref{Wdef} gives us an optimal joint probability distribution
function, $\pi_{ij}^{*}(k,l)$ for a set of distributions $m_i$ and $m_j$.
We define mixed states, $\hat\rho'(i,j)$ in $QDH$ by,
\begin{equation}
\hat\rho'(i,j)\equiv\sum_{k,l}\pi_{ij}^{*}(k,l)\vert k,l\rangle\langle k,l\vert.
\end{equation}

The squared Wasserstein distance $W^{(2)}(m_i,m_j)$
given by Eq. \ref{Wdef} can be rewritten as,
\begin{equation}
W^{(2)}(m_i,m_j) = 1 - Tr(\hat\rho_0\hat\rho').
\end{equation}

\subsubsection{Analytical and numerical results}
\label{res}
In this section we present the results obtained from the analytical calculations
with the known distances obtained from the ground state at the extreme limits of the interaction 
and the numerical results obtained for the distance matrices given by exact diagonalization, 
for interaction values $V=0-12$ and system sizes $L \leq 18$.

\begin{figure}[h!]
\includegraphics[width=0.21\textwidth]{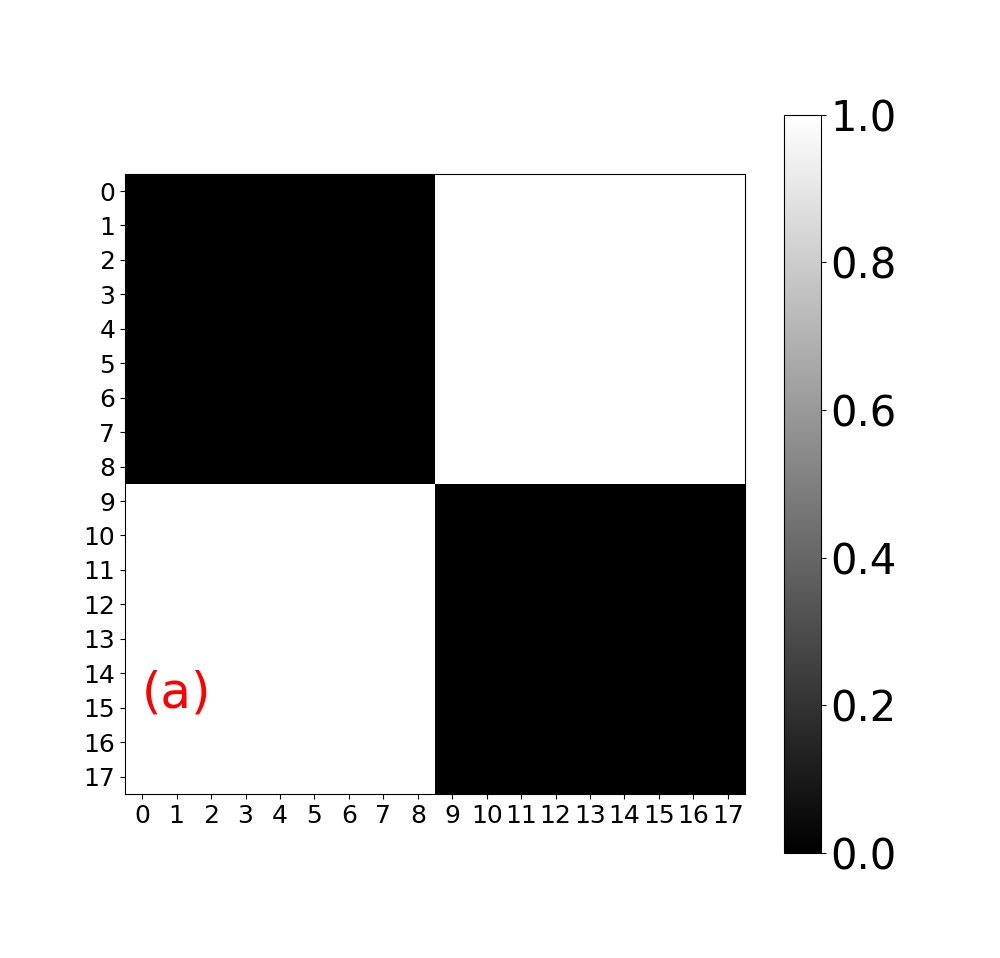}
\includegraphics[width=0.21\textwidth]{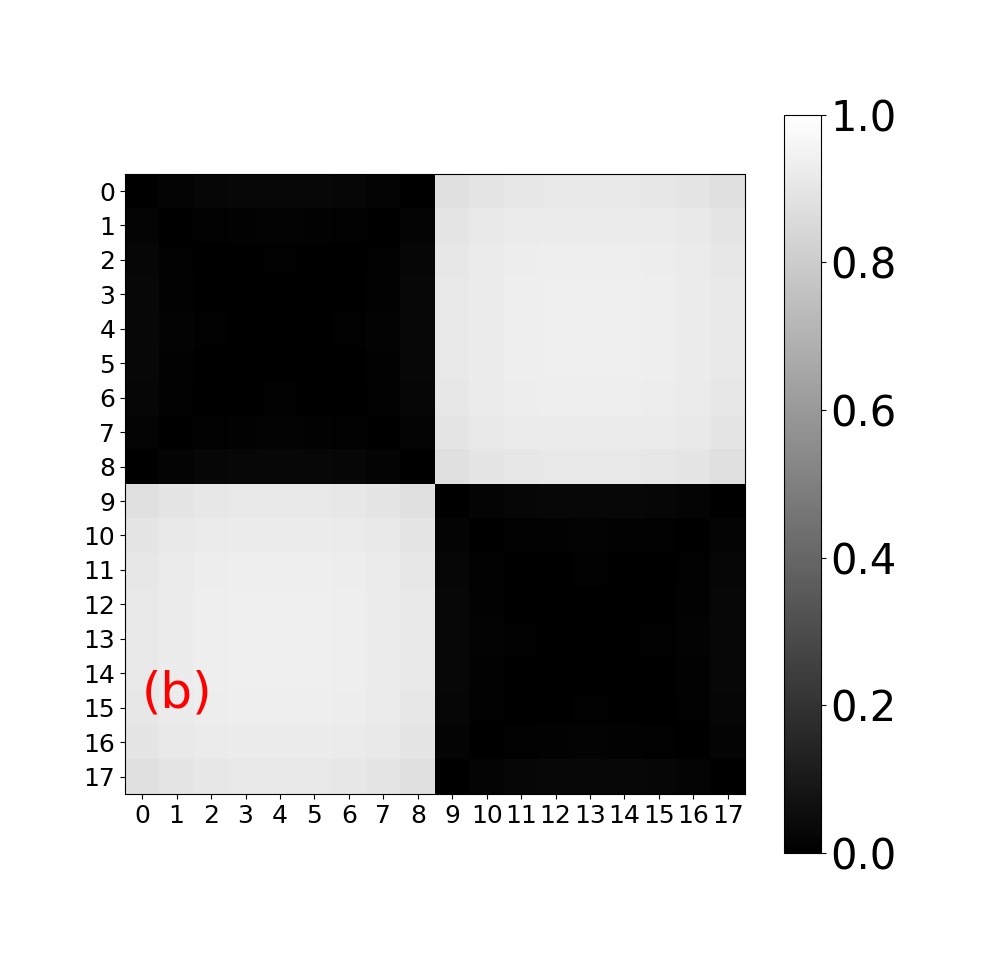}\\
\includegraphics[width=0.21\textwidth]{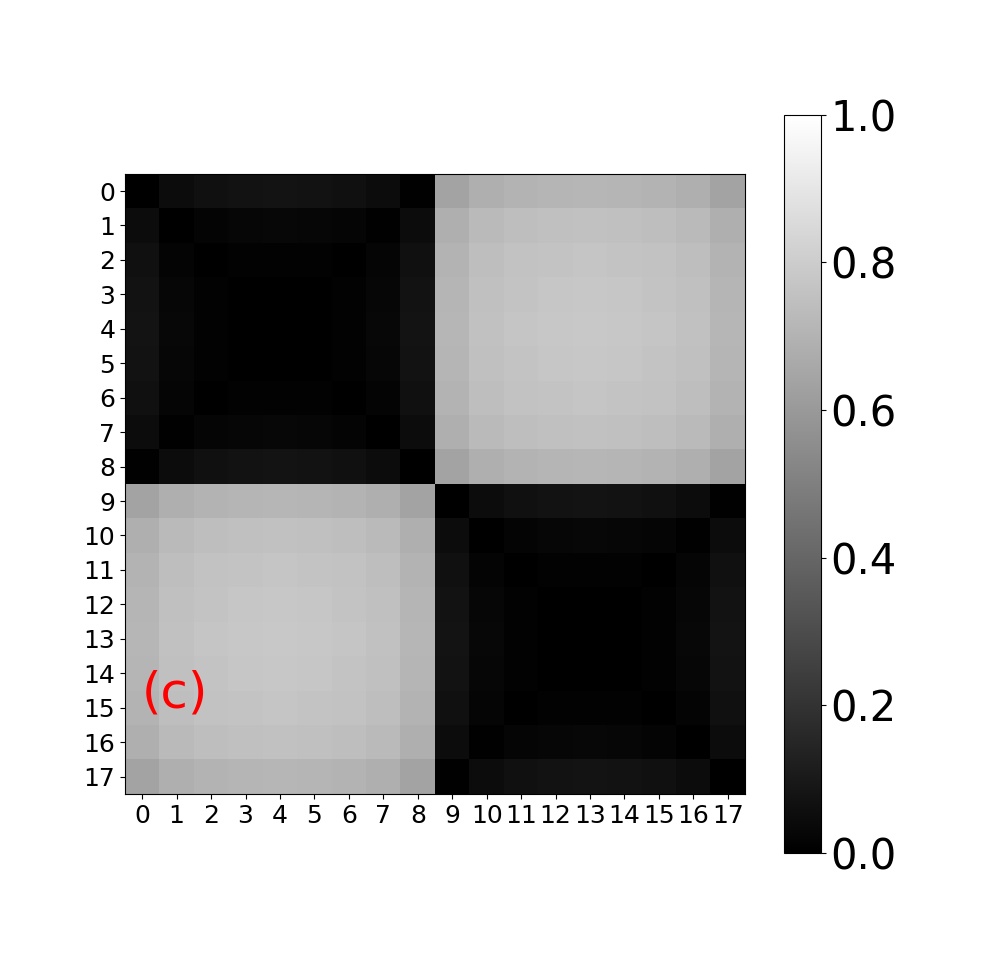}
\includegraphics[width=0.21\textwidth]{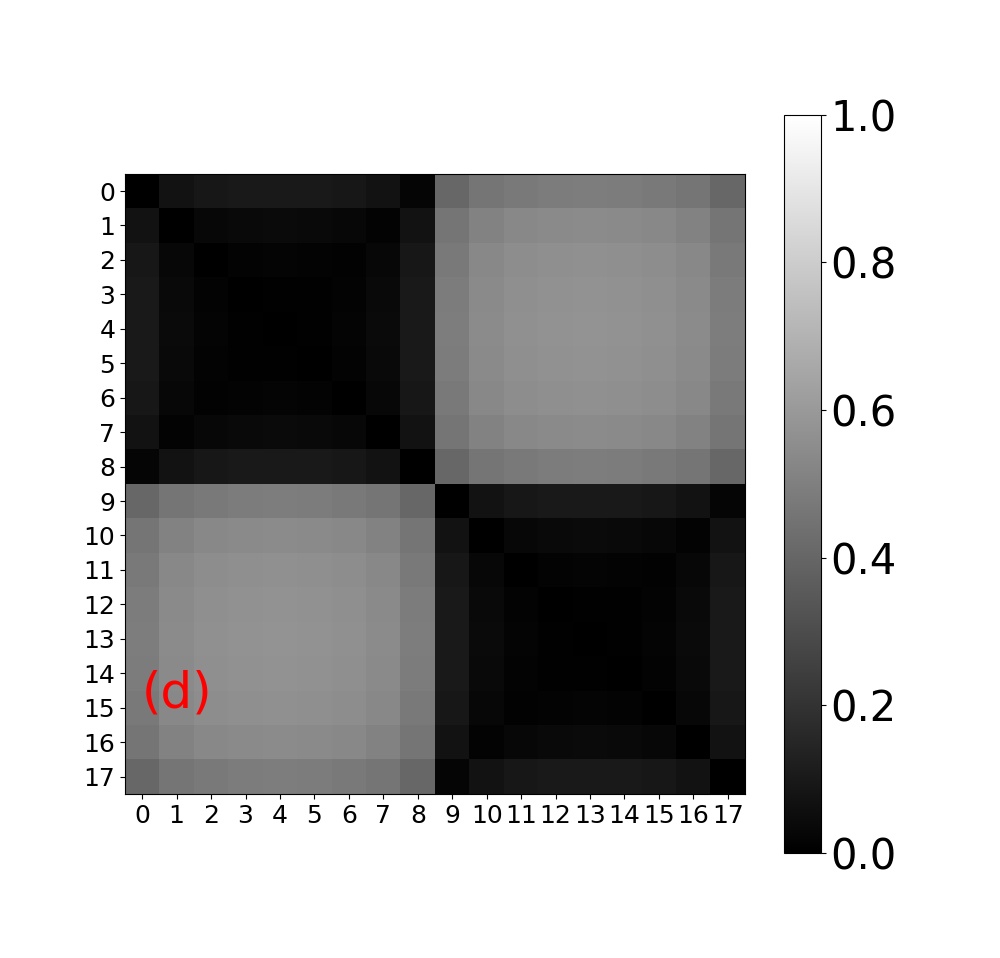}\\
\includegraphics[width=0.21\textwidth]{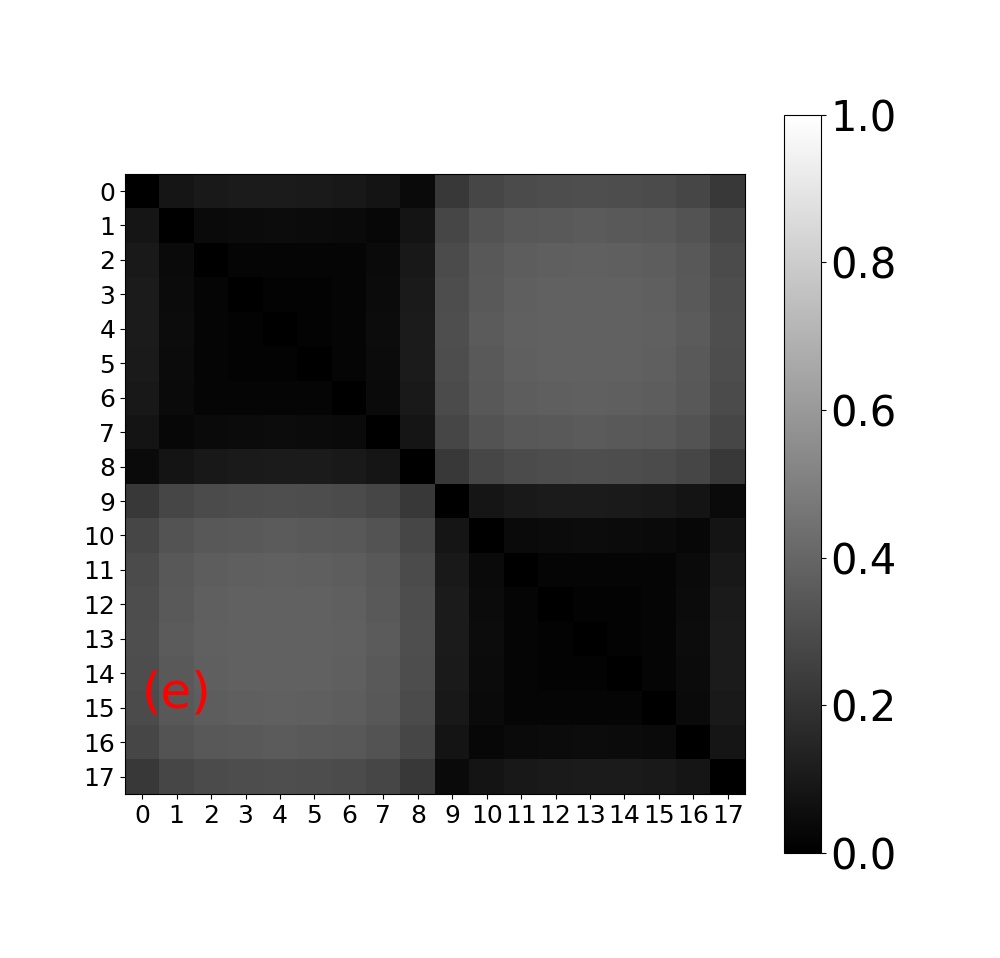}
\includegraphics[width=0.21\textwidth]{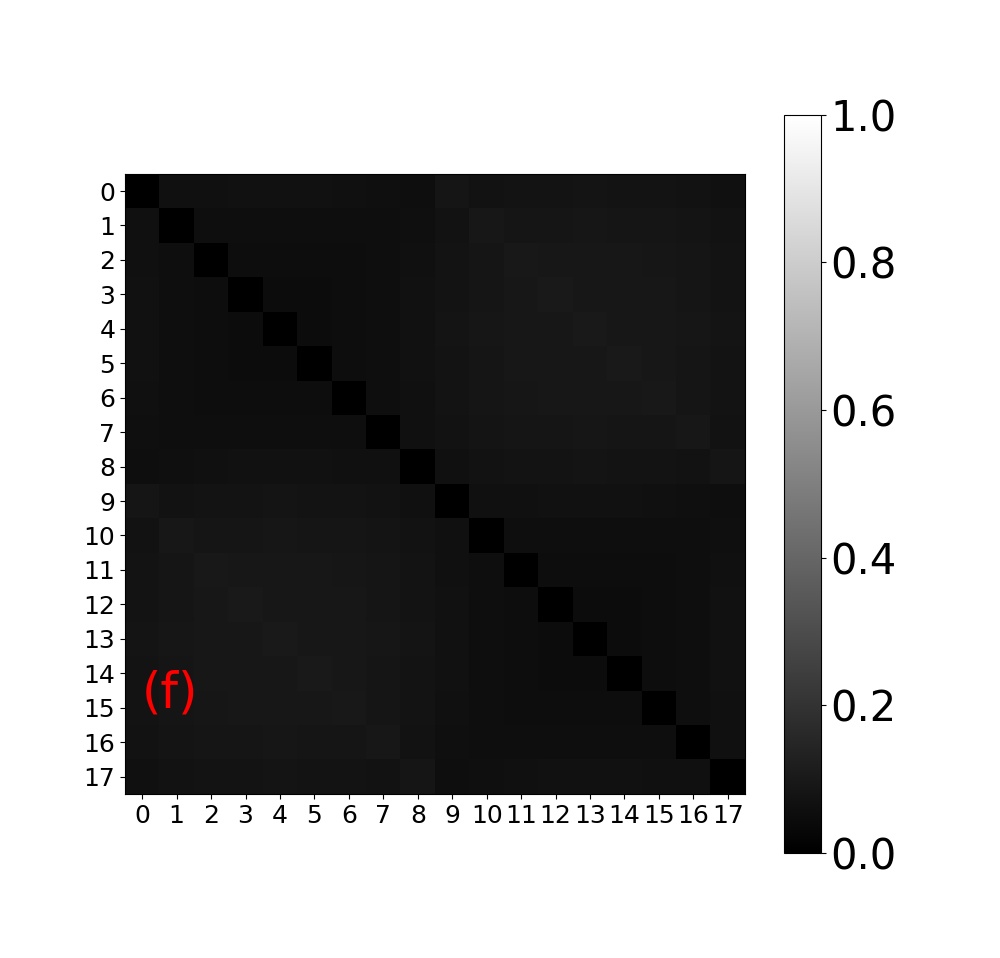}\\
\caption{(a)-(f)Squared Wasserstein Distance matrices $W^{(2)}(m_i,m_j)$ for $L=18$, obtained from 
numerical computation for interaction strengths $V=0.1$ (a), $V=1$ (b), $V=2$ (c), 
$V=3$ (d), $V=4$ (e) and $V=12$ (f). For $i < 9$, the quasi momenta modes lie inside the Fermi sea,
$k_i \in k_{in}$ and for $i \geq 9$, the quasi-momenta modes lie outside it, $k_i \in k_{out}$. The deep 
metallic regime is characterised by $\sim 0, \sim 1$ distance values between $k_{in}-k_{in}$ ($k_{out}-k_{out}$) 
and $k_{in}-k_{out}$ blocks respectively. The deep insulating regime is characterised by uniform extremely 
small values very close to zero.}
\label{Matrix}
\end{figure}

We denote the analytical quantum distances in the ground state at extreme values
of coupling constant $V$ by $D_{V}(i,j)$ and the corresponding squared Wasserstein distances by 
$W^{(2)}_{V}(m_i,m_j)$. For the extreme limits of interaction we can then show that 
(calculations in \ref{Appendix})
\begin{eqnarray}
\label{Wlim}
W^{(2)}_{0}(m_i,m_j)&=&(D_{0}(i,j))^{2}\nonumber\\
W^{(2)}_{\infty}(m_i,m_j)&=&\frac{1}{{\cal L}}(D_{\infty}(i,j))^{2}.
\label{W_large}
\end{eqnarray}

Starting from the $L$ distribution functions defined on the BZ, for a system with $L$ number of lattice sites, 
comparing every pair of distributions ($m_i,~m_j$) we have a $L \times L$ matrix of squared 
Wasserstein distances $W^{(2)}(m_i,m_j)$.
We look at the numerical $W^{(2)}$ matrices obtained for interaction values $0<V \leq 12$ in Fig. (\ref{Matrix}).
We find a direct reflection of the features of the distribution functions observed in Sec.~\ref{DDF}.
In deep metallic regime, the distance distributions for $k_{in}-k_{in}$ or $k_{out}-k_{out}$ being identical
corresponding Wasserstein distances $W^{(2)}(m_{k_{in}},m_{k_{in}})$ (or $W^{(2)}(m_{k_{out}},m_{k_{out}})$) 
are very small ($\sim 0$). While, distributions for $k_{in}-k_{out}$ are completely opposite to each other
and thus the Wasserstein distances $W^{(2)}(m_{k_{in}},m_{k_{out}})$ are very large ($\sim 1$).
Whereas in deep insulating regime the distance distributions are homogenous and almost identical so the
Wasserstein distances are uniform and almost zero. 

\begin{figure}[h!]
\includegraphics[angle=0,width=0.45\textwidth]{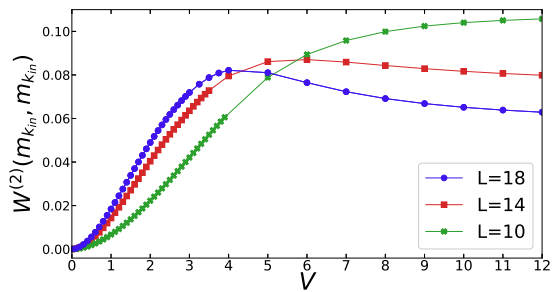}
\caption{Squared Wasserstein distances between quasi-momenta modes inside the Fermi sea,
$W^{(2)}(m_{k_{in}},m_{k_{in}})$, as a function of the interaction strength $V$
for system sizes $L=10,14,18$.}
\label{fig2}
\end{figure}

\begin{figure}[h!]
\includegraphics[angle=0,width=0.45\textwidth]{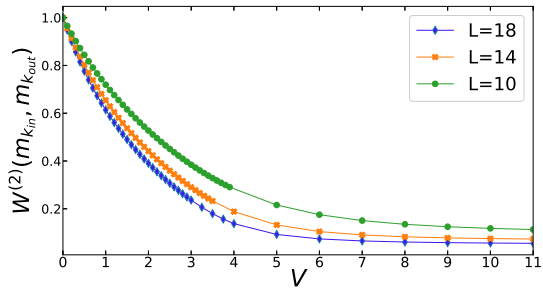}
\caption{Squared Wasserstein distances between quasi-momenta modes inside the Fermi sea and those
outside it, $W^{(2)}(m_{k_{in}},m_{k_{out}})$, as a function of the interaction strength $V$
for system sizes $L=10,14,18$.}
\label{fig3}
\end{figure}

We look at the behaviour of the distances $W^{(2)}(m_{k_{in}},m_{k_{in}})$ and
$W^{(2)}(m_{k_{in}},m_{k_{out}})$, as a function of the interaction strength $V$ 
for different system sizes in Fig.~(\ref{fig2}) and Fig.~(\ref{fig3}) respectively.
From Fig. (\ref{fig2}) we can expect 
$W^{(2)}(m_{k_{in}},m_{k_{out}})$ to indicate the critical interaction strength   
for the Luttinger liquid to CDW transition \cite{Shankar}, by the occurence of a
peak in the thermodynamic limit.

\begin{figure}[h!]
\includegraphics[angle=0,width=0.45\textwidth]{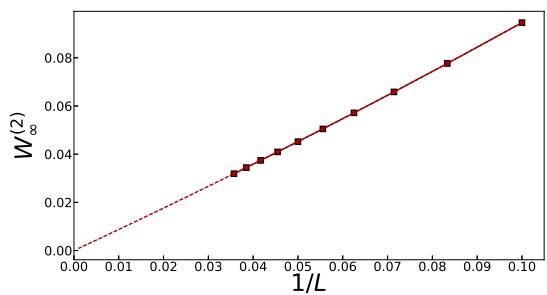}
\caption{$W_{\infty}^{(2)}$ as a function of the inverse system size $L^{-1}$
for system sizes $L=10-28$. It is found to be linear in $L^{-1}$ and thus vanishes in the
thermodynamic limit.}
\label{fig4}
\end{figure}

We numerically compute the Wasserstein distances in an ideal CDW phase 
starting from the distance distributions as given by the analytical distance matrices 
at $V= \infty$ (Eq.~\ref{dinfmat}) for system sizes $L=10-28$.
We denote the above uniform squared distances by $W_{\infty}^{(2)}$ and study its behaviour
as a function of the inverse of system size in Fig. \ref{fig4}. It is found to be linear in $L^{-1}$
and thus vanishes in the thermodynamic limit, as also predicted by Eq. \ref{W_large}.
The same result is indicated by Fig.~(\ref{fig2}) and Fig.~(\ref{fig3}) for $V\succsim 4$, in the 
insulating phase. While Eq.~(\ref{W_large}) and Fig.~(\ref{fig3})
suggest that the Wasserstein distances are non-zero for the metallic phase, in the 
thermodynamic limit.

From the above results we conclude, the Wasserstein distance gives a vivid geometric description 
of the ground state in both the phases. It characterises the geometry of the distance distributions 
extremely well. It characterises the metallic phase by classifying the points inside the Fermi sea 
and those outside it in two different groups, while it characterises the insulating phase by homogenous values.
The most striking finding is that in the thermodynamic limit,
the Wasserstein distance becomes zero in the insulating phase while it is non-zero in the metallic phase. 
Thus it provides a sharp characterisation of the phases of the system.

\subsection{The Wasserstien distance obtained from the Euclidean distance}
\label{WED}
In this section we propose a definition for Wasserstein distances in terms of the Euclidean distances
between the quasi-momenta in the BZ, by choosing the square of the Euclidean distances 
as the cost function of the optimal transport problem and substituting these distances in Eq.~(\ref{Wdef_gen}). 
We then look at the numerical results and observations.

We define the squared Wasserstein distance between two distance distributions $m_i$ and $m_j$, 
$W_{E}^{(2)}(m_i, m_j)$, as follows:

\begin{equation}
W_{E}^{(2)}(m_i, m_j) \equiv \inf_{\pi} \sum_{k,l=1} ^{L} (k-l)^{2} \pi_{ij} (k, l),
\label{WEdef}
\end{equation}
where $\pi_{ij}$ satisfies the constraints given by Eq. \ref{constraints}.

The above distance compares the distance distributions and equips us with means to study
the geometry of the distance distributions and thus the rich physical transformation emerging 
as a function of the interaction (Sec. \ref{DDF}). However, we cannot connect it to quantities
in the Hilbert space immediately as the distance function is no longer derived from states in
the Hilbert space. It can neither be connected to the intrinsic curvature as before. 
But it is still interesting physically, because it probes the geometry of the 
distance distribution functions derived from the quantum distances.

\subsubsection{Numerical results}
\label{res2}
\begin{figure}[h!]
\includegraphics[width=0.21\textwidth]{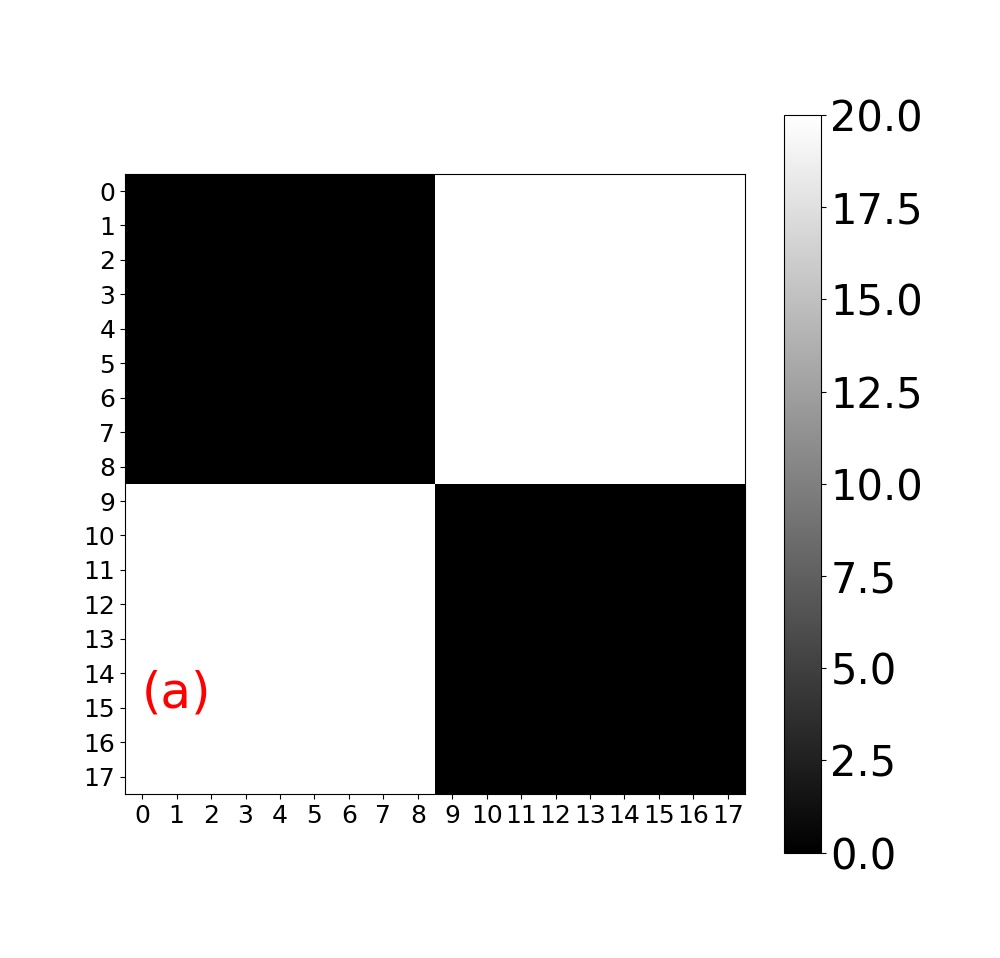}
\includegraphics[width=0.21\textwidth]{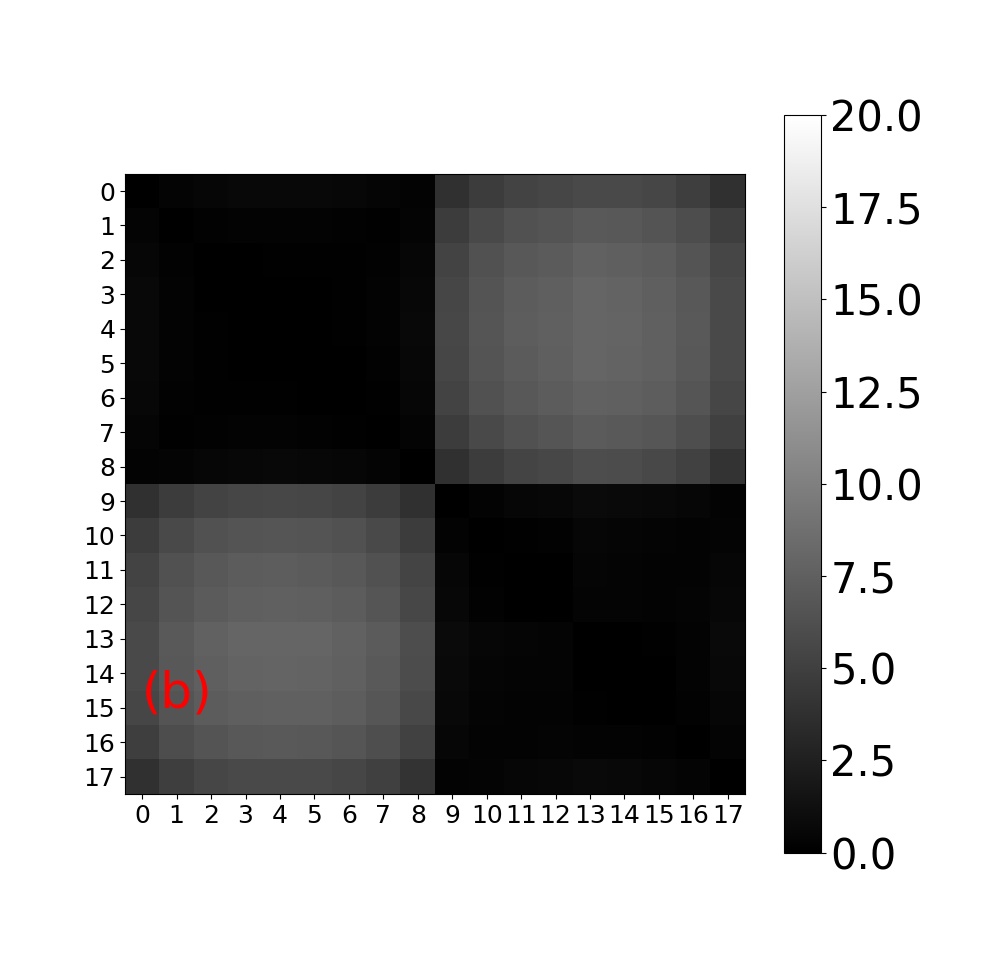}\\
\includegraphics[width=0.21\textwidth]{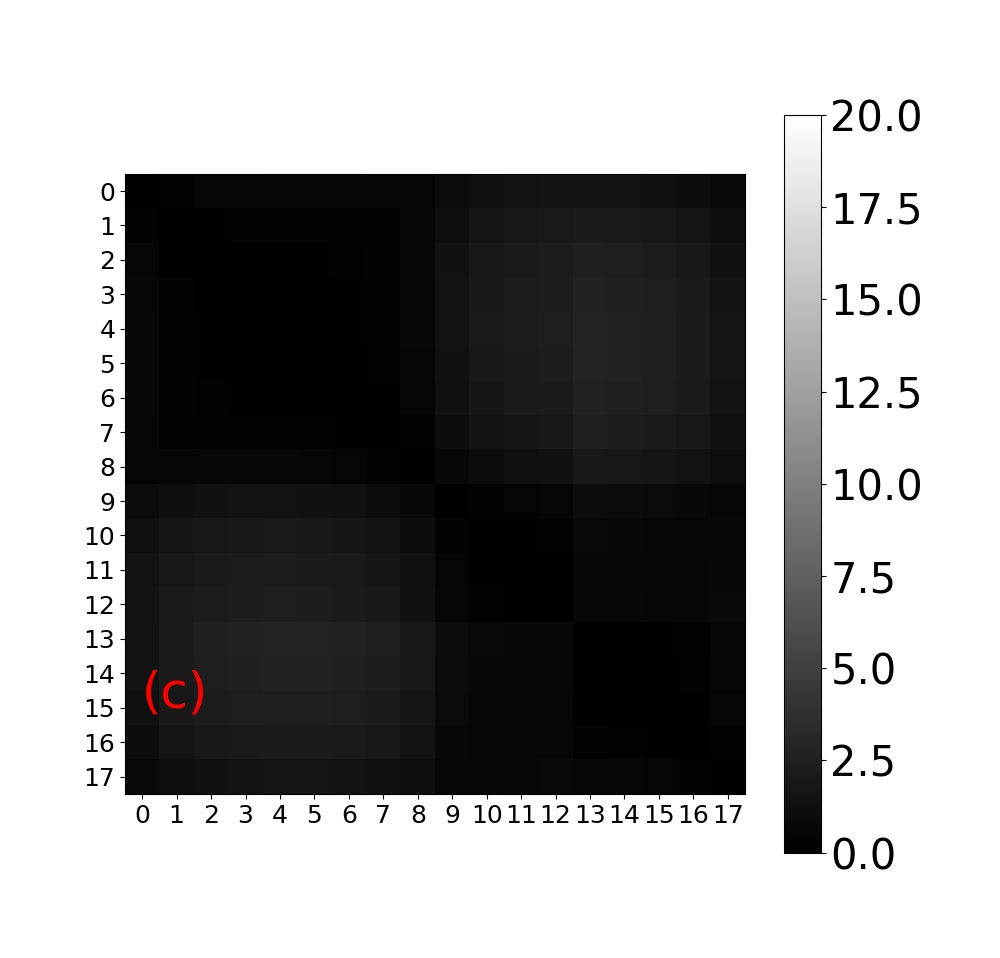}
\includegraphics[width=0.21\textwidth]{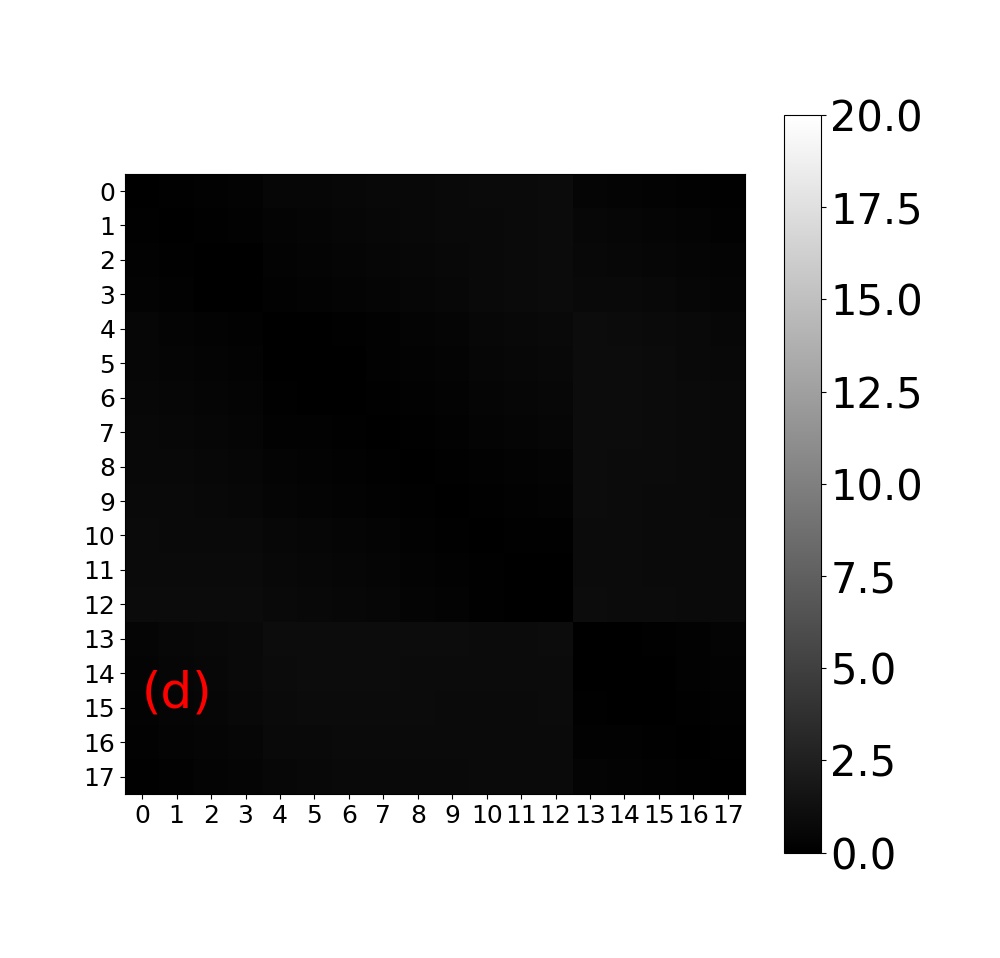}\\
\caption{(a)-(d)Squared Wasserstein Distance matrices $W_{E}^{(2)}(m_i,m_j)$ for $L=18$, obtained from 
numerical computation for interaction strengths $V=0.1$ (a), $V=2$ (b), $V=4$ (c), 
and $V=12$ (d). For $i < 9$, the quasi momenta modes lie inside the Fermi sea,
$k_i \in k_{in}$ and for $i \geq 9$, the quasi-momenta modes lie outside it, $k_i \in k_{out}$. The deep 
metallic regime is characterised by $\sim 0, \sim 20$ distance values between $k_{in}-k_{in}$ ($k_{out}-k_{out}$) 
and $k_{in}-k_{out}$ blocks respectively. The deep insulating regime is characterised by uniform and extremely 
small values.}
\label{Matrix_2}
\end{figure}
In this section we discuss the results obtained numerically from the linear programming solutions of 
the optimal transport problem, choosing the marginals to be the distribution functions constructed from 
the distance matrices obtained by performing exact diagonalization.  

We look at the numerical $W_{E}^{(2)}$ matrices obtained for interaction values $0< V \leq 12$ in Fig. (\ref{Matrix_2}).
We find the overall behaviour is quite similar to the observations found
for the Wasserstein distances obtained from the quantum distances in Sec. \ref{WQD}.
Like before, in deep metallic regime the Wasserstein distances classify the points inside and outside the Fermi
sea into two different categories, reflecting the geometry of the distance distributions (Secs. \ref{DDF} and \ref{WQD}).
The Wasserstein distances $W_{E}^{(2)}(m_{k_{in}},m_{k_{in}})$ (or $W_{E}^{(2)}(m_{k_{out}},m_{k_{out}})$) 
are very small ($\sim 0$), while $W^{(2)}_{E}(m_{k_{in}},m_{k_{out}})$ are very large ($\sim20$).
Whereas, in deep insulating regime the Wasserstein distances are uniform and very small.

\begin{figure}[h!]
\includegraphics[angle=0,width=0.45\textwidth]{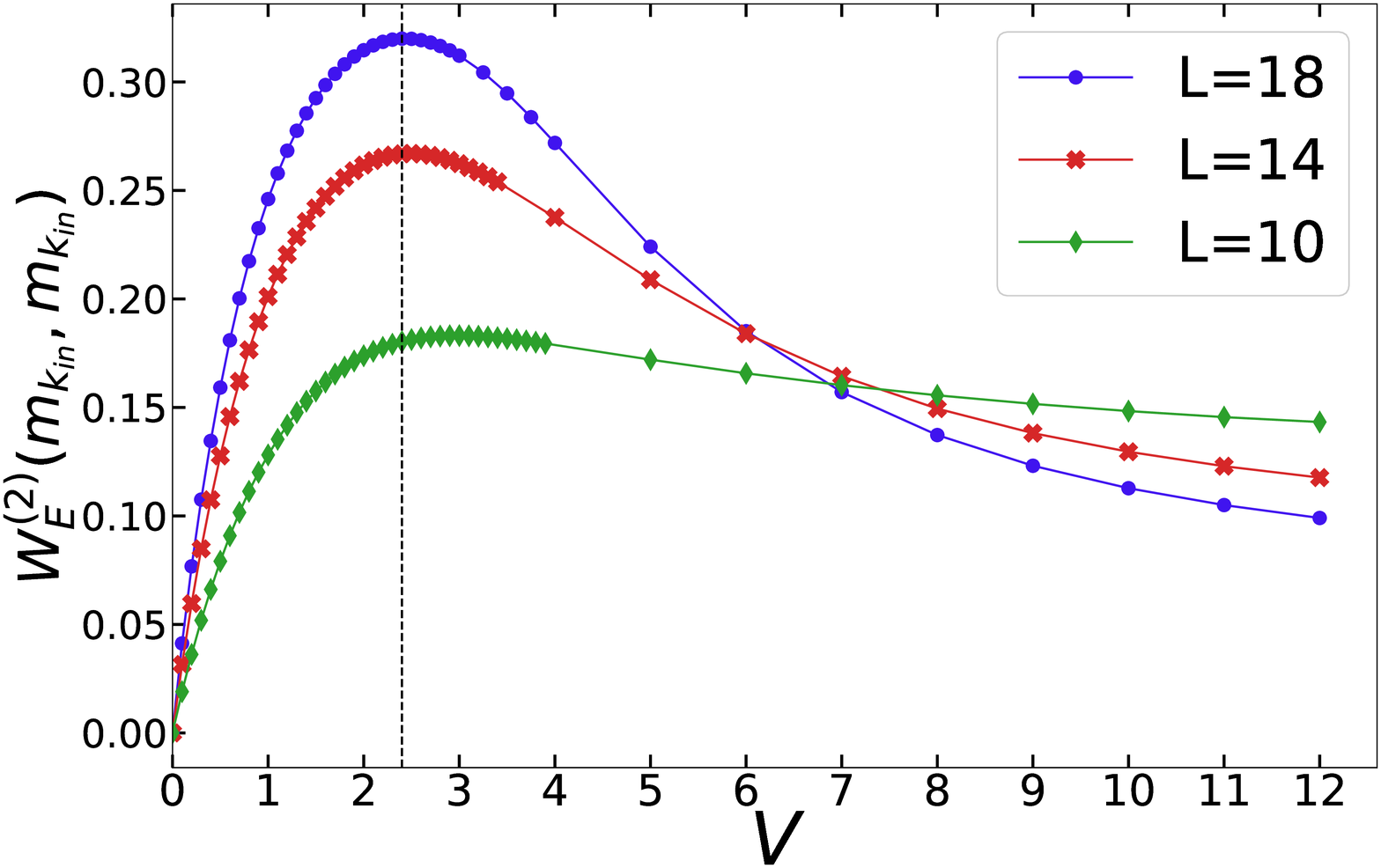}
\caption{Squared Wasserstein distances between quasi-momenta modes inside the Fermi sea,
$W_{E}^{(2)}(m_{k_{in}},m_{k_{in}})$, as a function of the interaction strength $V$
for system sizes $L=10,14,18$. The above distances show a prominent peak at $V=2.4$, for $L>10$.}
\label{fig2E}
\end{figure}

\begin{figure}[h!]
\includegraphics[angle=0,width=0.45\textwidth]{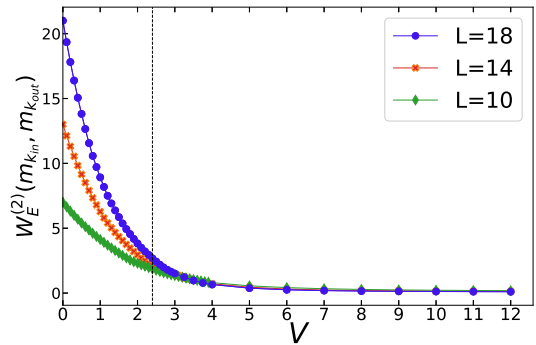}
\caption{Squared Wasserstein distance between quasi-momenta modes inside the Fermi sea and those
outside it, $W_{E}^{(2)}(m_{k_{in}},m_{k_{out}})$, as a function of the interaction strength $V$
for system sizes $L=10,14,18$. For $V \precsim 2.4$, above distance scales proportionately 
with the system size, post which it is insensitive to the system size.}
\label{fig3E}
\end{figure}

We look at the behaviour of the distances $W_{E}^{(2)}(m_{k_{in}},m_{k_{in}})$ and
$W_{E}^{(2)}(m_{k_{in}},m_{k_{out}})$, as a function of the interaction strength $V$ 
for different system sizes in Fig.~(\ref{fig2E}) and Fig.~(\ref{fig3E}) respectively.
The above distances sense the metal-insulator transition very distinctively and indicate
a critical interaction strength $V_c =2.4$, which is very close to the theoretical
value \cite{Shankar}. $W_{E}^{(2)}(m_{k_{in}},m_{k_{in}})$ indicates the transition by a peak,
which we expect to be more sharpened in the thermodynamic limit.
In the regime $V \precsim 2.4$, $W_{E}^{(2)}(m_{k_{in}},m_{k_{out}})$ rapidly reduces with the increase of 
interaction, moreover it scales proportionately with the system size, post which it is insensitive to the 
system size and constant as a function of the interaction. Thus $W_{E}^{(2)}(m_{k_{in}},m_{k_{out}})$ is 
expected to diverge in the metallic phase and remain finite in the insulating phase.

Summarising the results we can say that the Wasserstein distance obtained from the Euclidean distances 
in the BZ is able to characterise the geometry of the distance distributions extremely well and thus
provides a detailed geometric characterisation of the geometry of the ground state in both the phases.
It is able to indicate a critical value of interaction $V_c=2.4$ for the metal-insulator
transition, which is very close to the theoretical value \cite{Shankar}, $V=2$.
Moreover it is able to differentiate the metallic and insulating phases because the
distances between the quasi-momenta modes inside the Fermi sea and those outside it are divergent in
the metallic phase and finite in the insulating phase.

\section{The Wasserstein barycenter}
\label{WBC}
In the previous sections starting from $L$ distance distributions $\{m_i(k)\}$ at each point $k_i~(i=1,...L)$ on the BZ
we studied the Wasserstein distances defined between every pair of distibutions and found it very efficiently captures 
the physics of the system. However, the $L \times L$ matrix of Wasserstein distances have $L_{C_{2}}$ independent elements. 
Thus we further face the question, how to identify a single order parameter which can identify the metallic and insulating phases well. 
We address this question in this section by applying the concept of Wasserstein barycenter \cite{WBC1,WBC2}. 

In geometry, for a configuration of points the barycenter usually implies the arithmetic mean of the coordinates.
We begin with the question that analogously for above configuration of $L$ distance distributions whether there is 
some way of obtaining a single average distribution on the BZ which can efficiently characterise the configuration.
In Euclidean case \cite{BCO}, for a collection of points $(x_1,...,x_p)$, the barycenter $x^{*}$,  is obtained by minimising the function 
$\sum_{i=1}^{p} \lambda_{i} \mid x - x_i \mid ^{2}$, where $\lambda_{i} \in [0,1]$ and $\sum_{i=1}^{p} \lambda_{i}=1$ . 

Optimal transport generalizes the same concept to a collection of probability distributions by 
considering the weighted sum of squared Wasserstein distances instead of the above weighted sum of 
the squared Euclidean distances and introducing a single distribution function which minimizes 
the above sum \cite{WBC1}. The barycenter $m^{*}(k)$ is defined as a single function on the BZ 
such that the average sum of the squared Wasserstein distances between the function and each of the distributions 
(sum defined on the RHS of Eq.~\ref{BC}), is minimised.
We define a single geometric parameter, the average squared Wasserstein distance between the barycenter
and the distance distributions, as the follows,
\begin{equation}
\label{BC}
J(m^{*})  \equiv \inf_{m}\frac{1}{L} \sum_{i=1}^{L} W^{(2)}(m_{i},m), 
\end{equation}
where we take all the weights to be uniform for simplicity.
For computation of the Wasserstein barycenter we use the entropic regularisation of the optimal transport problem \cite{Cuturi}.
In above method, the solution for the barycenter is obtained by minimising the average sum of squared regularised Wasserstein distances
$W^{(2)}_{\gamma}(m_{i},m)$, defined by the following equation,
\begin{eqnarray}
\label{BCdef}
 W^{(2)}_{\gamma}(m_{i},m) &\equiv&  \inf_{\pi} (\sum_{ij} (D(i,j))^{2}\pi^{i}_{\gamma}(i,j) + \gamma S(\pi^{i}_{\gamma}))\\
 S(\pi^{i}_{\gamma}) &\equiv& \sum_{kl}\pi^{i}_{\gamma}(k,l) log(\pi^{i}_{\gamma}(k,l)).
\end{eqnarray}
 Where $\pi^{i}_{\gamma}(i,j)$ are the joint probability distributions with marginals $m_{i}$ and $m$, $\gamma$ is a positive
 regularisation parameter which is also a measure of the error introduced in the Wasserstein distance
 and $D(i,j)$ is the matrix of the quantum distances.
  The second term on the RHS corresponds to the entropy of above $\pi^{i}_{\gamma}(i,j)$ matrix. The regularised Wasserstein distances,
 can be computed by applying Sinkhorn-Knopp's fixed point iteration algorithm \cite{sinkhorn1967,Knight}.

\subsection{Numerical Results}
\label{res3} 
In this section we look at the numerical results obtained by taking the $18$ distance distribution functions constructed
from the quantum distances and computing the corresponding barycenter, defined in terms 
of the squared regularised Wasserstein distances $W^{(2)}_{\gamma}(m_{i},m)$ given by Eq.~(\ref{BCdef}), for a choice of the 
regularising parameter $\gamma=0.006$.

\begin{figure}[h!]
\includegraphics[angle=0,width=0.45\textwidth]{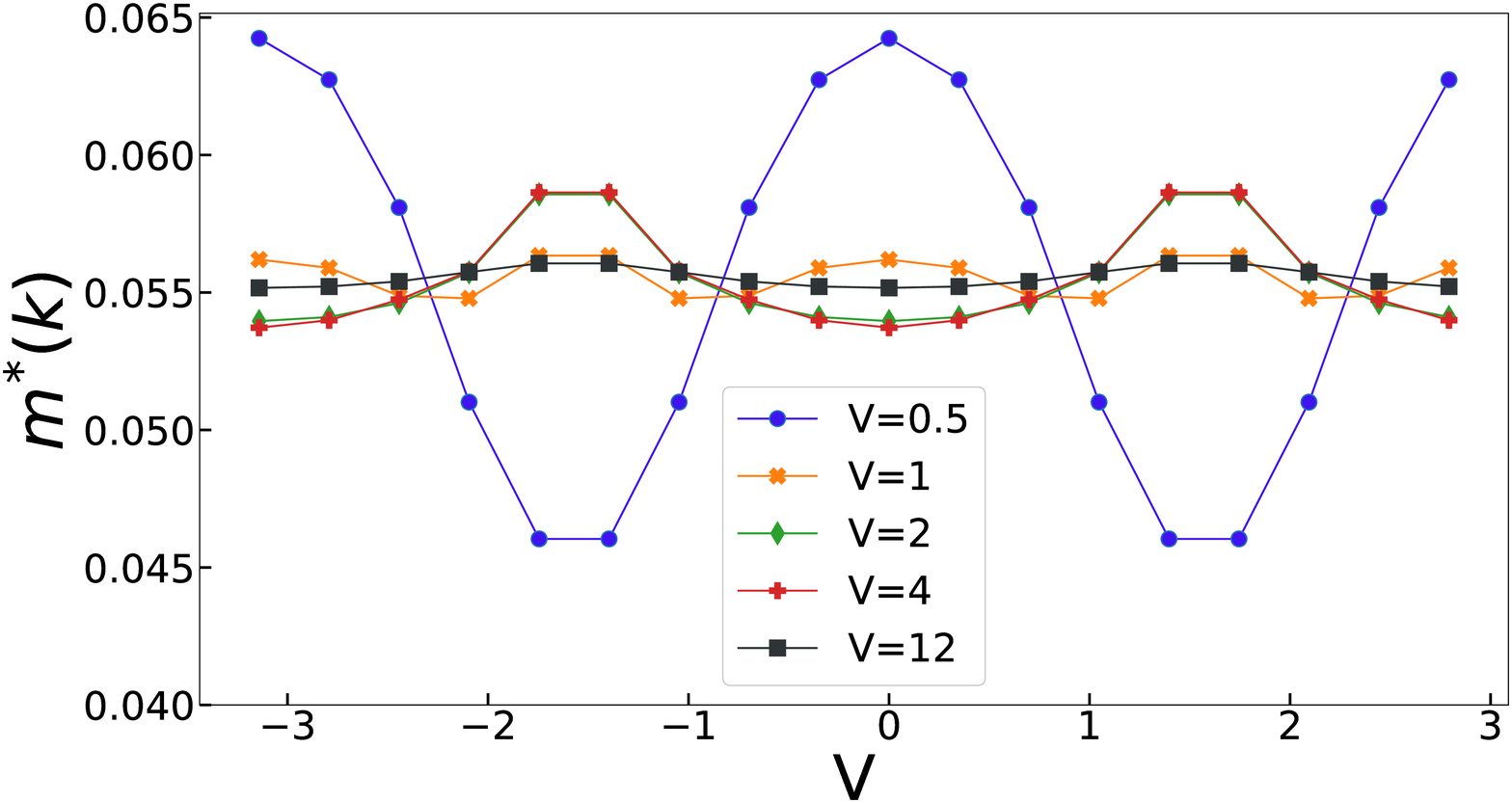}
\caption{The barycenter $m^{*}(k)$ defined over the BZ, $k \in [-\pi,\pi)$, for different interaction
values. The Fermi points are $k_f= \pm \frac{\pi}{2}$. We find that in deep metallic phase $m^{*}(k)$
is highly inhomogenous over the BZ and $m^{*}(k_f)$ is minimum. 
While in the deep insulating phase the distribution is flat and homogenous.}
\label{fig1_BC}
\end{figure}
We plot the barycenter over the BZ, for different interaction values in Fig.(\ref{fig1_BC}).
The distribution is highly inhomogeneous over the quasi-momenta modes in the metallic phase. 
For large $V$, in deep insulating phase we observe a contrasting homogenous behaviour.

In the  entropic regularisation method of the optimal transport, as discussed earlier, minimisation of the RHS of Eq.~(\ref{BCdef}), 
gives us an optimal but approximate joint distribution $\pi^{i*}_{\gamma}(k,l)$ which however is very close to the actual
$\pi^{i*}(k,l)$ for the extremely small value of regularisation parameter we choose, $\gamma=0.006$.
Thus average of the squared quantam distances over the BZ given by above joint probability distribution $\pi^{*i}_{\gamma}(k,l)$, 
will be very close to the squared Wasserstein distances  $W^{(2)}(m_{i},m)$ given by Eq.~\ref{Wdef}. So after obtaining the 
optimal joint distribution $\pi^{i*}_{\gamma}(k,l)$ by applying Sinkhorn-Knopp's fixed point iteration algorithm 
\cite{sinkhorn1967,Knight}, we redefine the corresponding squared
regularised Wasserstein distances $\tilde{W}^{(2)}_{\gamma}(m_{i},m^{*})$, as follows:
\begin{equation}
\label{WRdef}
 \tilde{W}^{(2)}_{\gamma}(m_{i},m^{*}) \equiv  \sum_{kl} (D(k,l))^{2}\pi^{i*}_{\gamma}(k,l).
\end{equation} 
The corresponding average squared Wasserstein distance between the barycenter and the configuration of $18$ distributions,
$\tilde{J}(m^{*})$ can then be defined as,
\begin{equation}
\label{Jdef}
 \tilde{J}(m^{*}) \equiv \frac{1}{L} \sum_{i} ( \sum_{kl} (D(k,l))^{2}\pi^{i*}_{\gamma}(k,l)).
\end{equation}

\begin{figure}[h!]
\includegraphics[angle=0,width=0.45\textwidth]{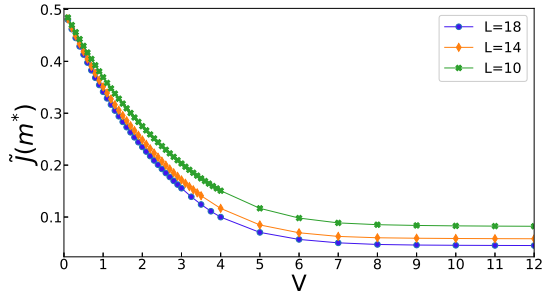}
\caption{Average squared Wasserstein distance between the distribution functions and the barycenter
as a function of the interaction strength, for different system sizes.}
\label{fig2_BC}
\end{figure}

\begin{figure}[h!]
\includegraphics[angle=0,width=0.45\textwidth]{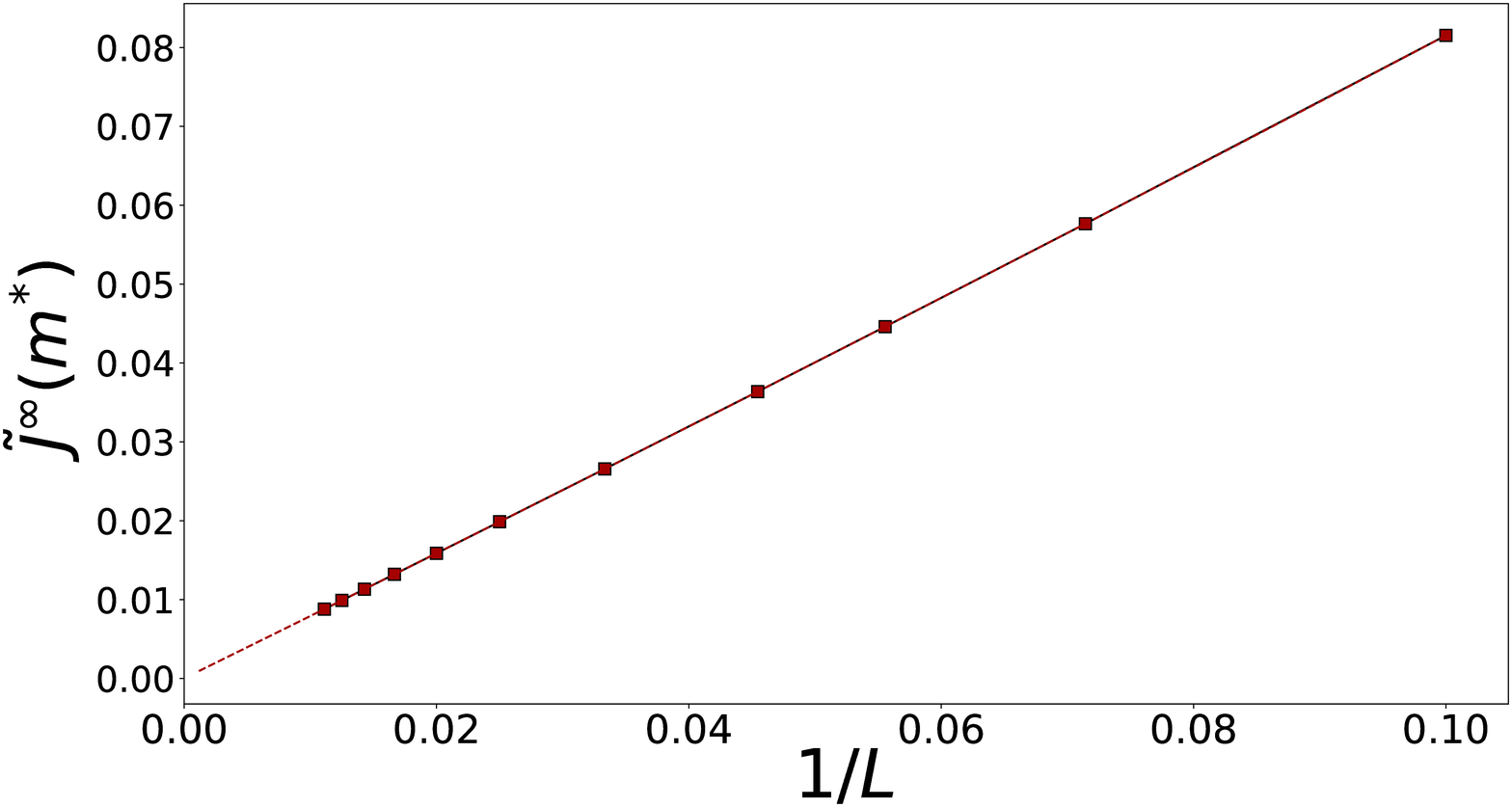}
\caption{Average squared Wasserstein distance between the distribution functions and the barycenter
at the extreme interaction limit, $V=\infty$ for system sizes $L=10-100$, as a function of the inverse of system size.}
\label{fig3_BC}
\end{figure}

We have looked at the behaviour of $\tilde{J}(m^{*})$ as a function of the interaction strength and 
compared system sizes $L=10,14,18$, like before, in Fig.~(\ref{fig2_BC}). We find it abruptly reduces
with increase of interaction strength in the metallic phase and becomes very small and rather insensitive
to the interacion in the deep insulating phase.
Moreover it is insensitive to the system size in deep metallic phase and reduces with increase of
system size in deep insulating phase. We compute $\tilde{J}(m^{*})$ for the CDW interacion
limit, $V=\infty$, by constructing the distributions from the analytical distance matrix (\ref{dinfmat}),
for system sizes $L=10-100$ and label it as $\tilde{J}^{\infty}(m^{*})$. It is found to be linear in $L^{-1}$
as demonstrated in Fig.~\ref{fig3_BC}, similar to the Wasserstein distances defined in terms of the
quantum distances in Sec.~\ref{WQD}.

In early metallic phase, with the distributions being drastically different
for points inside and outside the Fermi sea, we would expect the barycenter to be very different from the
starting parent distributions and corresponding average squared Wasserstein distance between the barycenter and the
distance distributions would be appreciable. However the insulating phase being characterised by more
or less identical distributions, the barycenter should also be an uniform distribution on the BZ
and the average Wasserstein distance from the parent distributions should be negligible in the 
thermodynamic limit. The above results provide strong evidences of the same.
The average Wasserstein distance between the barycenter and the distance distributions
becomes zero in the insulating phase and is non-zero in the metallic phase.
It can be considered as a single geometrical observable which is able to characterise the
phases.

\section{Discussion and conclusions}
\label{dandc}
As stated in the introduction (\ref{intro}), this paper along with 
two previous ones \cite{Paper1,Paper2} constitutes our attempt, following
previous work \cite{Walterkohn,RestaSorella,SCPMR,MarrazzoResta,resta2011} 
to formulate a new approach to the geometric 
characterization of insulating and metallic states. We first summarize
our results.

In our first work \cite{Paper1}, we had given a mathematically consistent
definition of the quantum distances between two points in the spectral
parameter space for a general many-fermion state. The quantum distances can be obtained 
by computation of the static correlation functions applying any exact or approximate technique 
like quantum Monte Carlo methods, DMRG and bosonisation in one dimension, 
exact diagonalisation for finite systems, perturbation theory etc..
We were motivated by previous work \cite{RestaSorella,SCPMR,MarrazzoResta,resta2011} 
which had related the concept of quantum distances to Kohn's seminal work 
\cite{Walterkohn} of understanding the structure of insulating states in 
terms of quantum geometry. Thus, we chose to implement and study our formalism 
in the 1-dimensional $t-V$ model, for reasons detailed in the introduction (\ref{intro}).
We found \cite{Paper1} that the distance matrix is qualitatively different 
in the metallic and insulating regimes.

In the following paper \cite{Paper2}, in an attempt to sharpen the
difference, we investigated the extrinsic and instinric geometry implied by the
distance matrix. We found that the intrinsic curvature, defined using the theory
of optimal transport \cite{OLLIVIER2009,OC2,ollivier:hal-00858008} and approximate Euclidean 
embedding of the Wasserstein distances seem to be the better discriminants .

Hence, in this paper we have focussed on applying the theory of optimal
transport \cite{Villani,villani2003} to analyse the quantum distance matrix of the 1 dimensional
$t-V$ model. We have obtained the following results.

We construct a geometric quantity, the matrix of Wasserstein
distances that distinguishes  sharply between the metallic and insulating
states, in the thermodynamic limit. The Wasserstein distances defined in terms
of the quantum distances are zero in the insulating phase and non-zero in the
metallic phase, while the Wasserstein distances defined in terms of the
Euclidean distances on the BZ, between the quasi-momenta modes inside the Fermi
sea and those outside it are divergent in the metallic phase and finite in the
insulating phase.

The matrices  of quantum distances and Wassertein distances are functions of a
pair of points in the spectral parameter space. It is clearly useful to
identify a single parameter to discriminate between the metallic and insulating
states. We have shown that the concept of the Wasserstein barycenter \cite{WBC1}, 
occuring in the theory of optimal transport, precisely does this. 

Thus, in the context of the one dimensional $t-V$ model, we have shown that the geometric
entities constructed using the theory of optimal transport give a sharp
distinction between the metallic and the insulating state.

It is therefore reasonable to infer that the theory of optimal transport,
in general is a good way to extract the physically relevant geometric
quantities of any correlated state. We would like to stress that it is not
our claim that by the procedure we adopted in these papers to analyse the metalllic
and insulating states we could observe a general characterisation for all 
physical situations. Each problem will have to be analysed individually. However, the general feature of this
technique, that it enables the extraction of geometric quantities averaged
over the BZ, lead us to propose that it will be useful in other physical
contexts as well. 

Our new approach to study the structure of a many-body state is closely related
with the machine learning approach towards data analysis. We generate
probability distribution functions from the many body state and by comparing these 
distributions ( in terms of Wasserstein distance) we infer geometric properties of
ground state. Similarly, in machine learning one often has to deal with
collections of samples that can be interpreted as probability distributions.
Comparing, summarising, and reducing the dimensionality of the probability
distributions on a given metric space are fundamental tasks in statistics
and machine learning. The Wasserstein distance is increasingly being used in
machine learning and statistics \cite{Cuturi,OT_2,OT_ML,OT_data,OT_ML_2}, 
especially for its way of comparing distributions based on basic principles.

In conclusion, we propose that, in general, physically relevant geometric 
information about correlated states can be extracted from the matrix of 
quantum distances using the techniques from the theory of optimal transport.

\section{Acknowledgements}
We are grateful to R.~Simon, S.~Ghosh, R.~Anishetty, A.~Samal and  
Emil Saucan for useful  discussions.

\appendix
\label{Appendix}
\section{The extreme limits}

In this section, we present analytic proofs for the results stated in the text
(Eq.~s \ref{Wlim}) for $W(m_i,m_j)$ in the extreme limits of the coupling.
Namely, $V=0$ and $V=\infty$.

\subsection{$V=0$}

The distance matrix at $V=0$ is easily computed \cite{Paper1}. It can
be written as
\begin{equation}
\label{d0mat}
D=\left(\begin{array}{cc}0&{\cal I}\\{\cal I}&0\end{array}\right)
\end{equation}
where ${\cal I}$ is the $L/2\times L/2$ matrix with all entries equal
to 1. We denote the $L/2$ component column vector with all entries equal
to 1 by $e$. We represent the distance distributions defined in Equation \ref{DF}
by column vectors $m_i,m_{L/2+i}, i=1,\dots,L/2$,
\begin{equation}
\label{mvecs}
m_i=\frac{2}{L}\left(\begin{array}{c}0\\e\end{array}\right)
~~~~~
m_{L/2+i}=\frac{2}{L}\left(\begin{array}{c}e\\0\end{array}\right).
\end{equation}
The constraints defining the joint probability distributions, $\pi_{ij}$
can be writen in a matrix form,
\begin{equation}
\label{picons}
\pi_{ij}\left(\begin{array}{c}e\\e\end{array}\right)=m_i
~~~~
\left(\begin{array}{cc}e^T&e^T\end{array}\right)\pi_{ij}=m^T_j.
\end{equation}
The general solution to the above equations(\ref{picons}) is
\begin{eqnarray}
\nonumber
\pi_{ij}=\frac{2}{L}\left(\begin{array}{cc}0&0\\0&P\end{array}\right)
~~~~
\pi_{iL/2+j}=\frac{2}{L}\left(\begin{array}{cc}0&0\\P&0\end{array}\right)\\
\label{pi0ans}
\pi_{L/2+ij}=\frac{2}{L}\left(\begin{array}{cc}0&P\\0&0\end{array}\right)
~~~~
\pi_{L/2+iL/2+j}=\frac{2}{L}\left(\begin{array}{cc}P&0\\0&0\end{array}\right),\nonumber
\end{eqnarray}
where $i,j=1,\dots,L/2$. $P$ is any $L/2\times L/2$ component positive 
semi-definite matrix whose rows and columns sum up to 1. $P(i,j)\ge 0,~
\sum_{k=1}^{L/2}P(k,l)=1=\sum_{l=1}^{L/2}P(k,l)$.

The Wasserstein distances can be written in this matrix form as,
\begin{equation}
\label{wmat}
W(m_i,m_j)=\inf_{\pi} {\rm Tr}\left(D^{(2)}\pi_{ij}\right),
\end{equation}
where $D^{(2)}(k,l)=(D(k,l))^{2}$.
Using the fact that $P{\cal I}={\cal I}={\cal I}P$, it is easy to see that 
the RHS of the above equation is independent of P and we obtain the result
stated in equation(\ref{Wlim}), that at $V=0$,
\begin{equation}
\label{0result}
W(m_i,m_j)=(D(i,j))^{2}.
\end{equation}

\subsection{$V=\infty$}

The distance matrix in the limit $V=\infty$ is \cite{Paper1},
\begin{equation}
\label{dinfmat}
D=c\left(
\begin{array}{cc}{\cal I}-I&{\cal I}\\{\cal I}&{\cal I}-I\end{array}\right)
+(1-c)\left(\begin{array}{cc}0&I\\I&0\end{array}\right),
\end{equation}

where $c=\sqrt{3}/2$. To represent the distance distributions as 
column vectors, we define a set of $L/2$ component column vectors,
$\chi_i,i= 1,\dots,L/2$ whose entries are all zero except for the $i^{\rm th}$ 
one, which is equal to 1. Namely, $\chi_i(j)=\delta_{ij}$. We then have,
\begin{equation}
\label{minfvec}
m_i=\frac{1}{\cal L}
\left(\begin{array}{c}ce-c\chi_i\\ce+(1-c)\chi_i\end{array}\right),
~~~~
m_{L/2+i}=\frac{1}{\cal L}
\left(\begin{array}{c}ce+(1-c)\chi_i\\ce-c\chi_i\end{array}\right)
\end{equation}
where ${\cal L}=c(L-2)+1$.
We present a set of solutions to equations. We have no proof that these
are the most general solutions. However, the definition of the Wasserstein
distance in equation(\ref{wmat}) implies that the infimum in this set 
provides an upper bound for $W(m_i,m_j)$. The solutions are of the form,
\begin{equation}
\label{piinfans1}
\pi_{ij}=\frac{1}{\cal L}\left(cP_{ij}+\pi^\prime_{ij}\right)
\end{equation}
where $P_{ij}$ are $L\times L$, positive semi-definite matrices whose
columns and rows sum up to 1. 
\begin{equation}
\label{pcons1}
P_{ij}\ge 0,~\sum_{k=1}^{L}P_{ij}(k,l)=1=\sum_{l=1}^{L}P_{ij}(k,l)
\end{equation}
$\pi^\prime_{ij}$ are,
\begin{eqnarray}
\nonumber
\pi^\prime_{ij}&=&
\left(\begin{array}{cc}-c\chi_i\chi^T_j&0\\0&(1-c)\chi_i\chi^T_j\end{array}
\right)\\
\nonumber
\pi^\prime_{iL/2+j}&=&
\left(\begin{array}{cc}0&-c\chi_i\chi^T_j\\(1-c)\chi_i\chi^T_j&0\end{array}
\right)
\\
\nonumber &&\\
\nonumber
\pi^\prime_{L/2+ij}&=&
\left(\begin{array}{cc}0&(1-c)\chi_i\chi^T_j\\-c\chi_i\chi^T_j&0\end{array}
\right)\\
\nonumber
\pi^\prime_{L/2+iL/2+j}&=&
\left(\begin{array}{cc}(1-c)\chi_i\chi^T_j&0\\0&-c\chi_i\chi^T_j\end{array}
\right)\\
\label{piinfans2}
\end{eqnarray}
Equations (\ref{piinfans1}), (\ref{piinfans2}) and the constraint
$\pi_{ij}(k,l)\ge 0, k,l=1,\dots,L$, implies that 
$ P_{ij}(i,j)-1\ge 0$. Since the maximum value of the matrix
elements of $P_{ij}$ is 1, we have,
\begin{equation}
\label{pcons2}
P_{ij}(i,j)=1,~~~~P_{ij}(i,k)=0~\forall k\ne j,~~~~P_{ij}(k,j)=0~\forall k\ne i.
\end{equation}
Consider the set of matrices, $P^*_{ij}$ defined as,
\begin{equation}
\label{pstardef}
P^*_{ij}(k,l)\equiv\delta_{kl}\left(1-\delta_{ik}-\delta_{jl}\right)
+\delta_{ik}\delta_{jl}+\delta_{il}\delta_{jk},
\end{equation}
it is straightforward to verify that $P^*_{ij}$ satisfy all the constraints
in equations (\ref{pcons1}) and (\ref{pcons2}).

Thus, equations (\ref{wmat}) and (\ref{piinfans1}) imply,
\begin{equation}
\label{wbound1}
W(m_i,m_j)\le \frac{1}{\cal L}
{\rm Tr}\left(D^{(2)}\left(cP^*_{ij}+\pi^\prime_{ij}\right)\right).
\end{equation}
The RHS of the above inequality can be computed using equations (\ref{dinfmat}),
(\ref{piinfans2}) and (\ref{pstardef}). The result is,
\begin{equation}
\label{wstarans}
{\rm Tr}\left(D^{(2)}\left(cP^*_{ij}+\pi^\prime_{ij}\right)\right)=(D(i,j))^{2}.
\end{equation}
Thus, we have proved the result stated in equation
\ref{Wlim}, that at $V=\infty$,
\begin{equation}
\label{infresult}
W(m_i,m_j)\le\frac{1}{\cal L}(D(i,j))^{2}.
\end{equation}

%

\end{document}